\documentclass[a4paper,12pt]{article}
\usepackage{amsmath,amssymb,amsthm}
\usepackage{cite}

\newtheorem{thm}{\bf Theorem}[section]

\newtheorem{prop}[thm]{\bf Proposition}

\numberwithin{equation}{section}

\newcommand{\al}{\alpha}
\newcommand{\be}{\beta}
\newcommand{\de}{\delta}
\let\ep\epsilon

\let\ph\phi
\def\tph{\tilde\ph}
\let\vph\varphi
\let\si\sigma
\let\th\theta
\let\z\zeta

\newcommand{\cH}{{\cal H}}
\newcommand{\cP}{{\cal P}}

\newcommand{\R}{{\mathbb R}}
\newcommand{\N}{{\mathbb N}}

\def\e{{\rm e}} \def\ii{{\rm i}}
\def\sL{\mathfrak{sl}}
\def\so{\mathfrak{so}}
\def\su{\mathfrak{su}}

\let\d\partial
\def\dz{\partial_z}
\def\dzz{\dz^2}
\let\l\ell
\def\abs#1{\left|#1\right|}

\let\Iff\Longleftrightarrow

\def\T{\tilde T}

\newcommand{\sn}{\operatorname{sn}}
\newcommand{\cn}{\operatorname{cn}}
\newcommand{\dn}{\operatorname{dn}}
\newcommand{\am}{\operatorname{am}}
\newcommand{\sx}{\sn x}
\newcommand{\cx}{\cn x}
\newcommand{\dx}{\dn x}
\newcommand{\dnx}{\dn^n x}
\newcommand{\cd}{\operatorname{cd} x}

\newcommand{\sd}{\operatorname{sd} x}
\newcommand{\sign}{\operatorname{sign}}
\newcommand{\Ec}{\operatorname{Ec}}
\newcommand{\Es}{\operatorname{Es}}
\renewcommand{\Re}{\operatorname{Re}}
\renewcommand{\Im}{\operatorname{Im}}

\def\tP{\tilde P}
\def\tp{\tilde p}
\def\htP{\hat{\!\tP}\!}
\def\tE{\tilde E}
\def\ta{\tilde a}
\def\tb{\tilde b}
\def\T{\tilde T}
\def\U{\tilde U}
\def\nt{\notag\\}

\begin{document}
    \title{\null\vskip-3cm\textbf{\Large A New Algebraization of the 
    Lam\'e Equation}%
\thanks{Supported in part by DGES Grant PB95--0401.}}
\author{\sc Federico Finkel\footnote{On
leave of absence from Depto.~de F\'\i sica Te\'orica II,
Univ.~Complutense de Madrid, Spain.}\\
\em Department of Mathematics\\
\em Imperial College\\
\em London SW7 2BZ, UK\\
\\
\sc Artemio Gonz\'alez-L\'opez, Miguel A.~Rodr\'\i guez\\
\em Departamento de F\'\i sica Te\'orica II\\
\em Universidad Complutense de Madrid\\
\em 28040 Madrid, SPAIN\\[.2cm]}
\date{July 29, 1999}
\maketitle
\begin{abstract}
We develop a new way of writing the Lam\'e Hamiltonian in
Lie-algebraic form. This yields, in a natural way, an explicit formula 
for both the Lam\'e polynomials and the classical non-meromorphic Lam\'e
functions in terms of Chebyshev polynomials and of a certain family of weakly 
orthogonal polynomials.
\end{abstract}
\newpage
\section{Introduction}
\label{sec.intro}
The Lam\'e equation,
\begin{equation}
    \label{lame}
    \psi''(x)+\left[E-m\,\l(\l+1)\,\sn^2 x\right]\psi(x)=0\,,
\end{equation}
where $\l$ is a real parameter\footnote{Note that without loss of generality
we can assume that $\l\geq-1/2$.}, and
$$
\sn x\equiv\sn(x|m)
$$
is the usual Jacobian elliptic function of modulus $m$, occupies a
central position in the theory of differential equations with periodic
coefficients. The study of its properties has attracted the attention
of many illustrious mathematicians over the last century; classical
references are~\cite{In39a, In39, EMOT55, Ar64}. Basic properties of
the Lam\'e equation are as follows. First, it arises by separation of
variables in the Laplace equation in ellipsoidal coordinates.
Secondly, it possesses two linearly independent $2K(k)$ or
$4K(k)$-periodic solutions (for characteristic values of $E$) if and
only if $\l$ is a nonnegative integer. Here $k=\sqrt m$, and $K(k)$
(denoted by $K$ from now on) is the complete elliptic integral of the
first kind with parameter $k$:
$$
K(k)=\int_0^{\pi/2}\frac{dx}{\sqrt{1-k^2\sin^2 x}}\,.
$$
Moreover, if $\l$ is a nonnegative integer there are exactly $\l$ gaps
in the energy spectrum, and the $2\l+1$ eigenfunctions associated to
the boundaries of the allowed energy bands can be written as
homogeneous polynomials of degree $\l$ in the elliptic functions
$\sn$, $\cn$, $\dn$. These polynomial solutions are known as the {\bf
Lam\'e polynomials}. In the third place, for {\em any} $\l\in\R$ and
characteristic values of $E$ the Lam\'e equation admits two linearly
independent solutions of period $8K$. These solutions are expressible
in closed form if and only if $\l\in\N+1/2$, and shall be referred to
as the {\bf non-meromorphic Lam\'e functions}. For both types of
solutions, namely the Lam\'e polynomials for integer $\l$ and the
non-meromorphic Lam\'e functions for half-integer $\l$, the
characteristic values of $E$ are the solutions of a certain algebraic
equation. We shall use the the term ``algebraic'' to refer to both
types of solutions.

The interest of the Lam\'e equation is by no means restricted to its
mathematical properties. In fact, several important physical
applications have been recently proposed in the literature. As first
remarked in~\cite{AGI83}, the Lam\'e potential can be considered as a
realistic model of a one-dimensional crystal. The Lam\'e equation also
describes some class of fluctuations of the sphalerons in the Abelian
Higgs model in 1+1 dimensions, \cite{BB93}. The quantum fluctuations
of the inflation field in certain cosmological models are determined
by the Lam\'e equation, \cite{KLS94, GKLS97}.

The Lam\'e equation appears in a natural way in several Lie-algebraic
approaches to the Schr\"odinger equation. Early work in this direction
was carried out by Alhassid, G\"ursey and Iachello in~\cite{AGI83}. In
this paper the Lam\'e equation was obtained by separation of variables
in the eigenvalue equation for a Hamiltonian quadratic in the
generators of $\su(2)$ written in conical coordinates. A different
Lie-algebraic representation of the Lam\'e equation was discovered by Turbiner
in~\cite{Tu89}, where it was shown that for integer $\l$ the Lam\'e
Hamiltonian belongs to the the enveloping algebra of a Lie algebra of
first-order differential operators with a finite-dimensional module of
functions. It follows that a number of eigenvalues and the
corresponding eigenfunctions of the Lam\'e equation can be
determined algebraically, i.e., by diagonalizing the finite-dimensional
matrix representing the action of the Hamiltonian in the module. This
class of Hamiltonians ---known in the literature as
{\em quasi-exactly solvable} (QES)--- have been the subject of
considerable investigation over the last decade; extensive reviews of
this field can be found in~\cite{GKOqes94, Us94, Ol97}. In particular,
it is known that several periodic potentials of physical interest
---like the Razavy potential,~\cite{Ra80, FGR99}, or the so-called
associated Lam\'e potential,~\cite{Ar64}--- possess the QES
property\footnote{A notorious exception is the Mathieu potential,
which is not QES.}.

In the above mentioned papers the Lam\'e equation was successfully
algebraized only for integer values of $\l$, corresponding to the
Lam\'e polynomials. A first indirect algebraization of the Lam\'e
equation for half-integer $\l$ was studied by Ward in~\cite{Wa87} in
connection with the matrix Nahm equations. The Lam\'e polynomials and
the non-meromorphic Lam\'e functions appear as solutions of some
matrix-valued differential equations related to irreducible
representations of $\so(3)$. The Lam\'e potential (and some
generalizations thereof) was revisited by Ulyanov and Zaslavskii
in~\cite{UZ92}. In this paper the Lam\'e Hamiltonian was obtained from
a non-standard realization of $\su(2)$ by first-order differential
operators, leading to the algebraic solutions for integer and
half-integer values of $\l$. In~\cite{BG93}, Brihaye and Godart
introduced yet another indirect algebraization scheme for half-integer
values of $\l$ based on a system of two second-order differential
equations.

In this paper we shall present a new direct way to algebraize the
Lam\'e equation which is valid for both the integer and half-integer cases.
Our approach is based on the classification of one-dimensional QES
potentials by Gonz\'alez-L\'opez, Kamran, and Olver, \cite{GKOqes94},
and presents some advantages over the similar algebraization considered
by Ulyanov and Zaslavskii in~\cite{UZ92}. Indeed, the $2\l+1$
algebraic points of the spectrum (corresponding to the boundaries of
the energy bands for integer $\l$ and to eigenfunctions of
quasi-momentum $\pi/(4K)$ for half-integer $\l$) are the roots of a
pair of polynomials of degrees $\l$ and $\l+1$ for integer $\l$, or of
a single polynomial of degree $\l+1/2$ for half-integer $\l$. This
compares to the algebraization by Ulyanov and Zaslavskii, for which
one needs to find the roots of a polynomial of degree $2\l+1$. In the
second place, it leads to new general expressions for the algebraic
solutions of the Lam\'e equation in terms of Chebyshev polynomials and
of a certain family of weakly-orthogonal polynomials. It should be
emphasized that in the algebraization schemes mentioned above explicit
solutions are presented only for low (integer or half-integer) values
of $\l$.

The paper is organized as follows. In Section 2, we introduce the
basic definitions and show how the Lam\'e potential can be obtained
using the classification of Gonz\'alez-L\'opez, Kamran, and Olver. The
algebraization by Ulyanov and Zaslavskii is briefly discussed in this
setting. Section 3 is devoted to the integer case. We find an
expansion in terms of Chebyshev polynomials which reveals the
structure of the classical Lam\'e polynomials. This expansion is used
to compute the Lam\'e polynomials for $\l\le4$. In the last
section we analyze the half-integer case. We obtain an
expansion of the non-meromorphic Lam\'e functions using again Chebyshev
polynomials, and relate our formulae to the classical results of Ince.
In particular, we check Ince's formulas for $\l\le 3/2$, and study in 
detail the case $\l=5/2$.
\section{Algebraizations of the Lam\'e equation}
\label{sec.alg}
If $n$ is a non-negative integer, it is a standard fact that the
differential operators
\begin{equation}
    \label{Js}
    J_-=\d_z\,,\qquad J_0=z\d_z-\frac n2\,,\qquad J_+=z^2\d_z-n\,z\,
\end{equation}
are the basis of a representation of the Lie algebra 
$\sL(2,\R)$ in the space $\cP_n$ of polynomials of degree at most $n$ 
in the variable $z$, \cite{Tu88, GKOnorm93}. As a consequence,
any differential operator $\cH$ which is a polynomial in the 
operators \eqref{Js} preserves the space $\cP_n$ (the converse is also 
true; cf.~\cite{Tu92, FK98}). In particular, $n+1$ 
eigenfunctions and eigenvalues of $\cH$ can be computed 
\emph{algebraically,} simply by diagonalizing the 
finite-dimensional operator obtained by restricting $\cH$ to $\cP_n$.
More generally, if we perform the change of variable
\begin{equation}
    z = \z(x)
    \label{cv}
\end{equation}
and consider the operator
\begin{equation}
    H(x)=\left.\mu(z)\cdot\cH\cdot\frac1{\mu(z)}\right|_{z=\z(x)}\,,
    \label{gauge}
\end{equation}
where $\mu(z)$ is a nowhere vanishing function, then $H$ obviously 
leaves invariant the $(n+1)$-dimensional vector space
$$
\mu\cP_n=\left\{\mu\bigl(\z(x)\bigr)\,p\bigl(\z(x)\bigr)\mid
p\in\cP_n\right\}\,.
$$
Consequently, we can again compute $n+1$ eigenvalues and
eigenfunctions of $H$ in an algebraic way, by considering its
restriction to the finite-dimensional space $\mu\cP_n$. We shall say
that such an operator $H$ is {\bf quasi-exactly solvable} (or, for
short, QES), \cite{TU87}. Of particular physical interest is the case
in which the operator $\cH$ is a polynomial of degree two in the
generators \eqref{Js}, which we shall write as
\begin{equation}
    \cH = -\sum_{a,b=0,\pm}c_{ab}\,J_a J_b-\sum_{a=0,\pm}c_a J_a-c_*\,.
    \label{HJ}
\end{equation}
Such an operator can be expressed as
\begin{equation}
    -\cH=P(z)\,\dzz+\left[Q(z)-\textstyle\frac12(n-1)\,P'(z)\right] \dz
    +R-\frac n2 Q'(z)+\frac 
    n{12}(n-1)P''(z)\,,
    \label{hg}
\end{equation}
where (see \cite{GKOnorm93})
\begin{align}
    P(z)&= 
    c_{++}\,z^4+2c_{+0}\,z^3+(2c_{+-}+c_{00})\,z^2+2c_{0-}z+c_{--}\,,\notag\\
    Q(z)&=c_+ z^2 + c_0 z + c_-\,,
    \label{PQR}\\
    R&=\frac n{12}(n+2)\,c_{00}-\frac n3(n+2)\,c_{+-}+c_*\,.\notag
\end{align}
Note that, using the Casimir identity
$$
J_0^2-\frac12\left(J_+J_-+J_-J_+\right)=\frac14 n(n+2)\,,
$$
we can take $c_{+-}=0$ without loss of generality. It can be shown,
\cite{GKOnorm93}, that if $P$ is positive then it is always possible
to transform (at least locally) the operator $\cH$ into a {\bf
Schr\"odinger operator} (or {\bf Hamiltonian})
\begin{equation}
    H = -\d_x^2+V(x)
    \label{schr}
\end{equation}
by an appropriate change of variable and {\bf gauge transformation}
\eqref{cv}--\eqref{gauge}. More precisely, the inverse of the change
of variable \eqref{cv} is given by
\begin{equation}
    x=\int^z_{z_0} \frac{dy}{\sqrt{P(y)}}\,,
    \label{cveq}
\end{equation}
while the gauge factor is proportional to
\begin{equation}
    \mu(z)=P(z)^{-n/4}\,\exp\left[\int^z_{z_0}\frac{Q(y)}{2P(y)}\,dy\right].
    \label{geq}
\end{equation}
The potential $V(x)$ can be expressed in terms of the 
coefficients of the {\bf gauge Hamiltonian} $\cH$ as follows
\begin{multline}
    V(x)=-\frac1{12\,P}\left[
    n(n+2)\left(P P''-\textstyle\frac34 P'{}^2\right)\right.\\
    \left.\left.{}+3(n+1)(QP'-2PQ')-3Q^2
    \right]\right|_{z=\z(x)}-R\,,
    \label{veq}
\end{multline}
where the prime denotes derivative with respect to $z$.

All one-dimensional potentials whose Hamiltonian is QES (in the
precise sense explained above) have been completely classified in
Ref.~\cite{GKOqes94}, modulo a constant translation of the space
coordinate $x$. In particular, the Lam\'e equation \eqref{lame} can be
obtained from the third family of periodic potentials (with $\nu=1$),
given by
\begin{equation}
    V(x)= A\,\sn^2 x+B\,\sx\,\cx+\frac{C\,\sx \cx+D}{\dn^2 x}\,.
    \label{case3}
\end{equation}
Note that, for later convenience, the above notation is slightly
different from that of Ref.~\cite{GKOqes94}. The potential 
\eqref{case3} is obtained from \eqref{veq} when
\begin{equation}
    P(z)=(1+z^2)\bigl[1+(1-m)z^2\bigr].
    \label{P}
\end{equation}
Indeed, from \eqref{cveq} it follows that, in this case,
\begin{equation}
    z = \frac{\sx}{\cx}\,.
    \label{zrule}
\end{equation}
Substituting \eqref{P} into \eqref{veq} and using \eqref{zrule} we
find the following explicit formula for the coefficients in
Eq.~\eqref{case3}
\begin{align}
    A&=\frac14 mn(n+2)-\frac{c_0}2(n+1) + 
    \frac1{4m}\bigr[c_0^2-(c_+-c_-)^2\bigr]\\
B&=\frac1{2m}(c_+-c_-)\bigl[m(n+1)-c_0\bigr]\\
C&=\frac1{2m}\bigl[c_+ +(m-1)c_-\bigr]\bigl[m(n+1)+c_0\bigr]\\
D&=\frac14(m-1)n(n+2)+\frac{c_0}{2m}(m-1)(n+1)\notag\\
&\kern9em{}+\frac1{4m^2}\left[(m-1)c_0^2+
\bigl(c_+ +(m-1)c_-\bigr)^2\right].
\end{align}
Note that we have taken
\begin{multline}\label{c*}
    c_*=\frac1{4m^2}\left[(2m-1)\,c_-^2+(1-m)\,(c_0^2+2c_+c_-)-c_+^2\right]
    \\
    +\frac{c_0}{2m}(n+1)-\frac14\,n(n+2)
\end{multline}
in order to eliminate a constant term in the potential. Comparing
\eqref{lame} with \eqref{case3} we obtain the system
$$
A=m\,\l(\l+1),\quad B=C=D=0\,,
$$
which has the following four sets of solutions:
\begin{alignat}{2}
    n&=\l;&\qquad c_+&=c_-=0,\quad c_0=-m\,\l\label{alg1}\\
    n&=\l-1;&c_+&=c_-=0,\quad c_0=-m\,(\l+1)\label{alg2}\\
    n&=\l-\frac12;& c_+&=c_-=\ii\,\sqrt{1-m},\quad 
    c_0=-m\,\left(\l+\textstyle\frac12\right)\label{alg3}\\
    n&=\l-\frac12;& c_+&=c_-=-\ii\,\sqrt{1-m},\quad 
    c_0=-m\,\left(\l+\textstyle\frac12\right).\label{alg4}
\end{alignat}
The first two solutions are valid when $\l$ is a non-negative integer
($\l\ge1$ for the second solution to exist), while the last two
solutions hold when $\l$ is a positive half-integer. From the previous
general discussion it follows that when $\l$ is a non-negative integer
or a positive half-integer a certain number of eigenfunctions and
eigenvalues of the Lam\'e Hamiltonian can be computed in a purely
algebraic fashion, by solving the eigenvalue problem of the
restriction of the operator $\cH$ corresponding to the Hamiltonian
\eqref{lame} to the finite-dimensional space $\cP_n$, where $n$ is
given by \eqref{alg1}--\eqref{alg4}. An explicit expression for the
gauge Hamiltonian $\cH$ can be easily found from Eqs.~\eqref{HJ},
\eqref{PQR}, \eqref{P}, and \eqref{alg1}--\eqref{alg4}, namely
\begin{equation}
    \label{hg14}
    -\cH = \left\{
     \begin{aligned}
    {}&(1-m)\,J_+^2+(2-m)\,J_0^2+J_-^2-m\,\l\,J_0+c_*\\
    {}&(1-m)\,J_+^2+(2-m)\,J_0^2+J_-^2-m\,(\l+1)\,J_0+c_*\\
    {}&(1-m)\,J_+^2+(2-m)\,J_0^2+J_-^2+\ii\,\sqrt{1-m}\,\left(J_+ 
    +J_-\right)\\{}&\kern15.5em{} -
    m\,\left(\l+\textstyle\frac12\right)\,J_0+c_*\\
    {}&(1-m)\,J_+^2+(2-m)\,J_0^2+J_-^2-\ii\,\sqrt{1-m}\,\left(J_+ 
    +J_-\right)\\{}&\kern15.5em{} -
    m\,\left(\l+\textstyle\frac12\right)\,J_0+c_*\,.
\end{aligned}
\right.
\end{equation}
Each of the four cases in this formula corresponds to the respective
solution \eqref{alg1}--\eqref{alg4}, and it is understood that in each
case the operators $J_\ep$ ($\ep=0,+$) and the constant $c_*$ must be
computed using the value of $n$ for the corresponding solution 
\eqref{alg1}--\eqref{alg4}. Note
that Eq.~\eqref{hg14} is essentially equivalent to
Eqs.~(7)--(9) of Ref.~\cite{Tu89}. The latter equations, however,
appear more complicated than \eqref{hg14}, due to the fact that 
Ref.~\cite{Tu89} uses the Weierstrassian form of Lam\'e's equation.

Before studying in more detail the four algebraizations 
\eqref{alg1}--\eqref{alg4} of the Lam\'e 
equation obtained in this section, it is worth pointing out that 
there are other alternative algebraizations within the general 
formalism described above. Indeed, from the identity
$$
\dn^{-2}(x+K)=\frac{\dn^2 x}{1-m}=\frac1{1-m}-\frac m{1-m}\,\sn^2 
x\,,
$$
and the fact that the potentials in Ref.~\cite{GKOqes94} are
classified up to constant translations, it can be shown that the
Lam\'e potential is a member of the first two families of periodic QES
potentials listed in the latter reference. For the same reason, we
could also have derived the Lam\'e equation by equating to zero the
coefficients $A$, $B$ and $C$ in Eq.~\eqref{case3}. It can be shown 
that all these ways of algebraizing the Lam\'e equation are essentially 
equivalent to the one adopted in this paper. Finally, there 
are other less obvious algebraizations of the Lam\'e equation like, 
for instance, the non-standard one discussed in Ref.~\cite{UZ92}, 
which falls into our framework by taking
$$
\cH = (1-m)\,J_0^2+\frac m4\left(J_+ + J_-\right)^2.
$$
From Eq.~\eqref{PQR} it follows that in this case
$$
P(z)=-\left[\frac m4\,z^4+\left(1-\frac m2\right)\,z^2+\frac m4\right],
    \quad Q(z)=0\,,\quad R=(2m-1)\,\frac{n(n+2)}{12}\,.
$$
It is straightforward to show that the \emph{complex} change of variable
$$
z=\cx+\ii\,\sx\,
$$
(cf.~Eq.~\eqref{cveq}) and the gauge transformation defined by
\eqref{gauge} and \eqref{geq}
map $\cH$ into the Schr\"odinger operator \eqref{schr} with potential
\begin{align}
    V(x)&=-m\,(1-m)\,\frac n2 \left(\frac 
    n2+1\right)\,\frac{\sn^2x}{\dn^2x}\nt
    &=m\,\frac n2 \left(\frac 
    n2+1\right)\,\left[\sn^2(x+K)-1\right]\,,
    \label{VUZ}
\end{align}
which is essentially the Lam\'e potential with $\l=n/2$. Since
$P(z)$ is negative everywhere and all its roots are complex, it can be
mapped to a negative multiple of the polynomial \eqref{P} by a
projective change of variable and gauge transformation. As explained
in Ref.~\cite{GKOqes94}, this means that the potential \eqref{VUZ}
can be obtained by a suitable choice of parameters in the third family
of QES periodic potentials (with pure imaginary frequency $\sqrt\nu$).
\section{Case I: $\l$ is a non-negative integer}
\label{sec.int}
We shall show in this section that when $\l$ is a non-negative integer the
algebraic eigenfunctions of the Lam\'e Hamiltonian are the
classical Lam\'e polynomials, \cite{Ar64}.

From Eqs.~\eqref{geq}, \eqref{zrule}, and
\eqref{alg1}--\eqref{alg2} we deduce that in this case the gauge
factor verifies
\begin{equation}
    \mu\left(\z(x)\right)=\begin{cases}
    \cn^\l x,& n=\l\\
    \cn^{\l-1} x\,\dx,& n=\l-1\,,\\
    \end{cases}
    \label{gl}
\end{equation}
where we have dropped some irrelevant numerical factors. From 
Eqs.~\eqref{gauge}, \eqref{zrule}, and \eqref{gl}, it follows that 
when $\l$ is a non-negative integer then the Lam\'e Hamiltonian has $\l+1$ 
eigenfunctions of the form
$$
    \psi(x) = \cn^\l x\cdot\chi\left(\frac{\sx}{\cx}\right)\,,
$$
where $\chi$ is a polynomial of degree at most $\l$. Thus we can write
\begin{equation}
    \psi(x) = \tilde\chi(\sx,\cx)\,,
    \label{sol1}
\end{equation}
where $\tilde\chi$ is a homogeneous bivariate polynomial of degree $\l$. 
Similarly, if $\l$ is 
strictly positive then there are $\l$ additional eigenfunctions of 
the Lam\'e Hamiltonian of the form
\begin{equation}
    \psi(x) = \dx\cdot\tilde\eta(\sx,\cx)\,,
    \label{sol2}
\end{equation}
where $\tilde\eta$ is a homogeneous polynomial in two variables of degree
$\l-1$. Thus, when $\l$ is a non-negative integer the Lam\'e 
Hamiltonian admits $2\l+1$ algebraic eigenfunctions of the form
\eqref{sol1}--\eqref{sol2}. 

To show that the latter $2\l+1$ algebraic
eigenfunctions are indeed the classical Lam\'e polynomials, we need to
study in more detail Eqs.~\eqref{sol1}--\eqref{sol2}. Fortunately,
this can be done in a totally systematic way, by applying the general
theory developed in Ref.~\cite{FGRorth96} for associating a (weakly)
orthogonal polynomial family to a one-dimensional QES Hamiltonian.
According to the prescription of the latter reference
(cf.~\cite{FGRorth96}, Eqs.~(30), (33), and (35)), we must choose two
different roots $z_1\ne z_2$ of $P(z)$ such that
\begin{equation}
    (2j-n-1)\,P'(z_1)+2\,Q(z_1)\ne0,\qquad\forall j\ge1\,.
    \label{cond}
\end{equation}
Then the function
\begin{equation}
    \psi_E(x)=\mu\bigl(\z(x)\bigr)\,\bigl(\z(x)-z_2\bigr)^n\,
    \sum_{j=0}^\infty 
    \frac{P_j(E)}{j!}\left(\frac{\z(x)-z_1}{\z(x)-z_2}\right)^j
    \label{psigen}
\end{equation}
(with $\z(x)$ and $\mu\bigl(\z(x)\bigr)$ respectively given by
Eqs.~\eqref{zrule} and \eqref{gl}), is a formal solution of the Lam\'e
equation with energy $E$ if and only if the set
$\left\{P_j(E)\right\}_{j\ge0}$ satisfies a certain $3$-term
recurrence relation. If we normalize $\psi_E$ so that $P_0=1$, the
latter recurrence relation implies that
$\left\{P_j(E)\right\}_{j\ge0}$ is a family of (weakly) orthogonal
polynomials. Moreover, if the coefficients of the recurrence relation
satisfy a positivity condition that we shall state and verify below,
the {\bf critical polynomial} $P_{n+1}$ has $n+1$ different real roots
$E_i$, $0\le i\le n$. Since it is easily shown that each of these
roots is also a root of the polynomials $P_j(E)$ with $j>n+1$, for
$i=0,1,\dots, n$ the function $\psi_{E_i}(x)$ is a genuine solution of
the Lam\'e equation \eqref{lame} with energy $E=E_i$. These $n+1$
exact solutions can be obtained \emph{algebraically,} by computing the
$n$ polynomials $P_1(E),\dots,P_n(E)$ from the recurrence relation,
and in fact coincide with the $n+1$ eigenfunctions of the Lam\'e
Hamiltonian $H$ algebraically computable by diagonalizing the
restriction of $H$ to $\mu \cP_n$ (or, equivalently, the restriction
of $\cH$ to $\cP_n$).

It is immediate to verify that \eqref{cond} is satisfied for both
solutions \eqref{alg1} and \eqref{alg2} simply by taking $z_1=-\ii$.
If we choose $z_2=\ii$, from Eq.~\eqref{psigen} we obtain the
following expression (up to a constant factor) for the $\l+1$ exact
eigenfunctions of the Lam\'e Hamiltonian coming from the first
algebraization \eqref{alg1}
\begin{equation}
    \label{psiei1}
    \psi_{E_i}(x)=\sum_{j=0}^\l
    \frac{(-1)^j}{j!}\,P_j(E_i)\,(\cx+\ii\sx)^{\l-j}(\cx-\ii\sx)^j\,,\qquad 
    0\le i\le \l\,,
\end{equation}
where the $\l+1$ energies $E_i$ satisfy
\begin{equation}
    \label{ei}
    P_{\l+1}(E_i)=0\,,\qquad 0\le i\le\l.
\end{equation}
In the same way, the second algebraization \eqref{alg2} yields the $\l$
eigenfunctions
\begin{multline}
    \label{psiei2}
        \psi_{\tE_i}(x)=\dx\,\sum_{j=0}^{\l-1}
    \frac{(-1)^j}{j!}\,\tP_j\,(\tE_i)(\cx+\ii\sx)^{\l-1-j}(\cx-\ii\sx)^j\,,\\
    \quad 0\le i\le \l-1\,,
\end{multline}
where $\left\{\tP_j\right\}_{j\ge0}$ is the orthogonal polynomial
family constructed from the solution \eqref{alg2}, and the $\l$
energies $\tE_i$ are given by
\begin{equation}
\label{eti}
\tP_{\l}(\tE_i)=0\,,\qquad 0\le i\le\l-1.
\end{equation}

We shall next derive the recurrence relation defining the polynomials
$P_j$ and $\tP_j$. According to the general procedure described in
Ref.~\cite{FGRorth96}, the latter relation is determined by the
coefficients of the auxiliary polynomials $\hat P(w)$, $\hat Q(w)$
obtained by applying to $P(z)$ and $Q(z)$ the projective
transformation
\begin{equation}
    w=\frac{z-z_1}{z-z_2}\,,
    \label{proj}
\end{equation}
namely
\begin{align}
    \hat
    P(w)&=\frac{(1-w)^4}{(z_1-z_2)^2}\,P\left(\frac{z_1-z_2\,w}{1-w}\right),
    \label{phat}\\ \hat
    Q(w)&=\frac{(1-w)^2}{z_1-z_2}\,Q\left(\frac{z_1-z_2\,w}{1-w}\right)
    \label{qhat}
\end{align}
(with $z_1=-\ii=-z_2$ in the present case). From Eqs.~\eqref{P} and
\eqref{phat} we easily obtain the following formula for $\hat P$, valid
for both algebraizations \eqref{alg1} and \eqref{alg2}
\begin{equation}
    \label{Phatint}
    \hat P(w) = m\,w^3 - 2\,(2 - m)\,w^2 + m\,w\,.
\end{equation}
Similarly, Eqs.~\eqref{PQR} and \eqref{qhat} imply that 
the polynomial $\hat Q$ associated to the first algebraization \eqref{alg1} 
is given by
\begin{equation}
    \label{qhat1}
    \hat Q = \frac12\,m\,\l\,(w^2-1)\,,
\end{equation}
while for the second algebraization \eqref{alg2} we obtain
\begin{equation}
    \label{qhat2int}
    \hat Q = \frac12\,m\,(\l+1)\,(w^2-1)\,.
\end{equation}
Following Ref.~\cite{FGRorth96},
we define new polynomials $\hat P_j(E)$ and $\htP_j(E)$ by
\begin{equation}
    \label{pphatint}
    \left.
    \begin{aligned}
	{}&\hat P_j(E)\\
	{}&\htP_j(E)
    \end{aligned}
   \right\}
    =\left(\frac m2\right)^j\frac{(2\l-1)!!}{(2\l-2j-1)!!}\,
    \left\{
    \begin{aligned}
	{}&P_j(E)\\
	{}&\tP_j(E)\,.
    \end{aligned}
    \right.
\end{equation}
It is shown in Ref.~\cite{FGRorth96} that the polynomials $\hat P_j$
satisfy the recurrence relation
\begin{equation}
    \hat P_{j+1}=(E-b_j)\,\hat P_j-a_j\,\hat P_{j-1}\,,\qquad j\ge 0\,,
    \label{rr}
\end{equation}
with $\hat P_{-1}=0$ and
\begin{align}
    a_j&=\frac14\,m^2\,j\,(2j-1)\,(2\l-2j+1)\,(\l-j+1)\,,
    \label{ajint}\\
    b_j&=\frac12\,m\,\l\,(\l+1)+\frac12\,(2-m)\,(\l-2j)^2\,.
    \label{bjint}
\end{align}
Likewise, the polynomials $\htP_j$ verify the recurrence relation
\begin{equation}
    \htP_{j+1}=(E-\tb_j)\,\htP_j-\ta_j\,\htP_{j-1}\,,\qquad j\ge 0\,,
    \label{rrt}
\end{equation}
where $\htP_{-1}=0$, and the coefficients $\ta_j$ and $\tb_j$ are given by
\begin{align}
    \ta_j&=\frac14\,m^2\,j\,(2j+1)\,(2\l-2j+1)\,(\l-j)\,,
    \label{taj}\\
    \tb_j&=\frac12\,m\,\l\,(\l+1)+\frac12\,(2-m)\,(\l-2j-1)^2\,.
    \label{tbj}
\end{align}
Note that from the normalization $P_0=\tP_0=1$ and Eq.~\eqref{pphatint} 
it follows that
\begin{equation}
    \label{normph}
    \hat P_0=\htP_0=1\,.
\end{equation}
Furthermore, Eqs.~\eqref{ajint} and \eqref{taj} imply that the
coefficients $a_j$ and $\ta_j$ are strictly positive for $1\le j\le
\l$ and $1\le j\le \l-1$, respectively. By Lemma 1 (Sec.~1.8) of 
Ref.~\cite{Ar64}, this positivity condition 
guarantees that all the roots of the critical polynomials $\hat 
P_{\l+1}$ and $\htP_\l$ are real and simple, as we had anticipated.

Our next task is to simplify Eqs.~\eqref{psiei1} and \eqref{psiei2} 
using the properties of the polynomials $P_j$ and $\tP_j$. We shall 
start by proving the following useful result: 
\begin{prop}\label{prop.reim}
    The algebraic eigenfunctions 
\eqref{psiei1} and \eqref{psiei2} are always either real or purely 
imaginary. 
\end{prop}
\begin{proof}
Let us rewrite 
Eqs.~\eqref{psiei1} and \eqref{psiei2} in terms of the elliptic 
amplitude $\vph(x)=\am x$, defined by
$$
\cx+\ii\,\sx=e^{\ii\,\vph(x)}\,.
$$
We thus obtain the expressions
\begin{align}
    \label{psiphi}
    \psi_{E_i}(x) &= \sum_{j=0}^\l p_{ij}\,\e^{\ii (\l-2j)\vph(x)},
    \qquad 0\le i\le \l\,,\\
    \intertext{and}
    \label{psitphi}
    \psi_{\tE_r}(x) &= \dx\,\sum_{s=0}^{\l-1}\tp_{rs}\,\e^{\ii 
    (\l-2s-1)\vph(x)},\qquad
    0\le r\le \l-1\,,
\end{align}
where we have set, for convenience,
\begin{equation}
    \label{pptij}
    p_{ij}=\frac{(-1)^j}{j!}\,P_j(E_i),\qquad
    \tp_{rs}=\frac{(-1)^s}{s!}\,\tP_s(E_r)\,.
\end{equation}
Let us concentrate, for definiteness, on the eigenfunctions of type 
\eqref{psiphi}. We rewrite 
Eq.~\eqref{psiphi} in the form
\begin{equation}
    \psi_{E_i}(x)=\sum_{0\le j<\l/2}\left[p_{ij}\,\e^{\ii (\l-2j)\vph(x)}
    +p_{i,\l-j}\,\e^{-\ii (\l-2j)\vph(x)}\right]+p_{i,\frac\l2},
    \label{psi2}
\end{equation}
where we take $p_{i,\l/2}=0$ when $\l$ is odd.
Taking into account that, by the recurrence relation 
\eqref{rr}--\eqref{bjint}, all the coefficients $p_{ij}$ 
are real, we conclude from the Eq.~\eqref{psi2} that
\begin{align}
    \label{repsi}
    \Re \psi_{E_i}(x)&=
    \sum_{0\le j<\l/2}\left(p_{ij}+p_{i,\l-j}\right)\,\cos(\l-2j)\vph(x)
    +p_{i,\frac\l2}\,,\\
    \Im \psi_{E_i}(x)&=
    \sum_{0\le 
    j<\l/2}\left(p_{ij}-p_{i,\l-j}\right)\,\sin(\l-2j)\vph(x)\,.
    \label{impsi}
\end{align}
Let us now introduce the polynomial family
\begin{equation}
    \pi_j(E)=\frac{(-1)^j}{j!}P_j(E),\quad
    \label{pipoly}
\end{equation}
in terms of which the coefficients $p_{ij}$ are simply expressed by
$$
p_{ij}=\pi_j(E_i)\,,\qquad 0\le i,j\le\l\,.
$$
Defining two additional families of univariate polynomials $\si_j(E)$ 
and $\rho_j(E)$ by
\begin{equation}
    \si_j(E)=\pi_j(E)+\pi_{\l-j}(E)\,,\quad
    \rho_j(E)=\pi_j(E)-\pi_{\l-j}(E)\,,
    \label{side}
\end{equation}
from Eqs.~\eqref{repsi}--\eqref{impsi} we have
\begin{align*}
\Re \psi_{E_i}\equiv0&\quad\Iff\quad \si_j(E_i)=0
\quad \text{for\quad} 0\le j\le[\l/2]\,,\\
\Im \psi_{E_i}\equiv0&\quad\Iff\quad \rho_j(E_i)=0
\quad \text{for\quad}  0\le j<\l/2
\end{align*}
(where $[{}\cdot{}]$ denotes the integer part). From
Eqs.~\eqref{pphatint}--\eqref{bjint} and \eqref{pipoly}, it
immediately follows that the polynomials $\pi_j(E)$ satisfy the
three-term recurrence relation
\begin{equation}
    (j+1)(2\l-2j-1)\pi_{j+1}=\frac2m(b_j-E)\pi_j-(2j-1)(\l-j+1)\pi_{j-1}
    \label{pirr}
\end{equation}
with the initial conditions
\begin{equation}
    \pi_{-1}(E)=0,\qquad\pi_0(E)=1\,,
    \label{pi-10}
\end{equation}
where the coefficients $b_j$ are defined by Eq.~\eqref{bjint}.
It is easy to see that Eq.~\eqref{side} implies that the polynomial families
$\left\{\si_j(E)\right\}_{j\ge0}$ and
$\left\{\rho_j(E)\right\}_{j\ge0}$ satisfy the {\it same} three-term
recurrence relation \eqref{pirr} as the family $\left\{\pi_j(E)\right\}_{j\ge0}$.
Moreover, from Eq.~\eqref{ei} we obtain
\begin{equation}
    \si_{-1}(E_i)=\rho_{-1}(E_i)=0\,,\qquad 0\le i\le\l\,.
    \label{minus1}
\end{equation}
From this, and the fact that the coefficients
$\left\{\si_j(E_i)\right\}_{j\ge0}$ and
$\left\{\rho_j(E_i)\right\}_{j\ge0}$ satisfy a \emph{three}-term recursion 
relation of the type \eqref{pirr}, we conclude that the vanishing of
$\si_0(E_i)$ or of $\rho_0(E_i)$ automatically implies the vanishing of 
$\si_j(E_i)$ or of $\rho_j(E_i)$, respectively, for all values of $j>0$. 
Thus we can write
\begin{equation}
    \Re \psi_{E_i}\equiv0\Iff \si_0(E_i)=0\,,\qquad
    \Im \psi_{E_i}\equiv0\Iff \rho_0(E_i)=0\,.
    \label{reim0}
\end{equation}
To complete the proof of the Proposition, we simply note that
Eqs.~\eqref{pi-10}--\eqref{minus1} and the fact that
$\left\{\si_j(E_i)\right\}_{j\ge0}$ and
$\left\{\rho_j(E_i)\right\}_{j\ge0}$ satisfy the same three-term
recursion relation as $\left\{\pi_j(E)\right\}_{j\ge0}$ imply that
$$
\si_j(E_i)=\si_0(E_i)\,\pi_j(E_i)\,,\qquad j\ge0\,,
$$
and, in particular
\begin{multline}
    \si_0(E_i)=\si_\l(E_i)=\si_0(E_i)\,\pi_\l(E_i)\\\Iff
    \si_0(E_i)\left[1-\pi_\l(E_i)\right]\equiv\si_0(E_i)\,\rho_0(E_i)=0\,.
    \label{sd0}
\end{multline}
The proof for the eigenfunctions of type \eqref{psitphi} 
is totally analogous, and will therefore be omitted.
\end{proof}

Let us order, from now on, the $\l+1$ algebraic eigenvalues 
$E_i$ of the Lam\'e equation in increasing order, 
i.e.,
$$
E_0<E_1<\dots<E_{\l-1}<E_\l\,,
$$
and similarly for the $\l$ eigenvalues $\tE_i$. From Lemma 1
(Sec.~1.8) of Ref.~\cite{Ar64} and the positivity of $a_j$ for $1\le
j\le\l$ we easily deduce that
\begin{equation}
    \sign\left[(-1)^\l P_\l(E_i)\right]=(-1)^i\,,
    \label{sign}
\end{equation}
which implies (by Eq.~\eqref{pipoly}) that $\pi_\l(E_i)$ has the sign 
of $(-1)^i$. Thus for even $i$ we have $\si_0(E_i)=1+\pi_\l(E_i)>0$, and 
from \eqref{sd0} we conclude that $\rho_0(E_i)=0$. In the same way we 
establish that $\si_0(E_i)=0$ for odd $i$. From Eq.~\eqref{reim0} we 
obtain
\begin{equation}
    \Re\psi_{E_{2r+1}}=\Im\psi_{E_{2r}}=0\,.
    \label{reimpsi0}
\end{equation}
We can therefore write the $\l+1$ exact eigenfunctions of type 
\eqref{psiei1} of the Lam\'e equation as the Jacobi--Fourier series
\begin{align}
    \label{lame1e}
    \ph_{2r}(x)&\equiv\psi_{E_{2r}}(x)=
    \sum_{0\le j<\l/2}\si_{2r,j} \cos(\l-2j)\vph(x)
    +p_{2r,\frac\l2}\\
    \ph_{2s+1}(x)&\equiv\psi_{E_{2s+1}}(x)=
    \sum_{0\le j<\l/2}\rho_{2s+1,j}\,\sin(\l-2j)\vph(x)\,,
    \label{lame1o}
\end{align}
with
\begin{equation}
    \si_{2r,j}\equiv\si_j(E_{2r})=p_{2r,j}+p_{2r,\l-j}\,,\quad
    \rho_{2s+1,j}\equiv\rho_j(E_{2s+1})=p_{2s+1,j}-p_{2s+1,\l-j}\,,
    \label{siders}
\end{equation}
and
\begin{equation}
    0\le r\le\left[\frac\l2\right],\qquad 0\le s\le\left[\frac{\l-1}2\right].
    \label{rangers}
\end{equation}
We can derive in a totally analogous way an equivalent formula for the 
$\l$ eigenfunctions of type \eqref{psi2}, namely
\begin{align}
    \label{lame2e}
    \tph_{2r}(x)&\equiv\psi_{\tE_{2r}}(x)=
    \dx\left[\tp_{2r,\frac12(\l-1)}+\sum_{0\le 
    j<\frac12(\l-1)}\tilde\si_{2r,j}\,\cos(\l-2j-1)\vph(x)\right],\\
\tph_{2s+1}(x)&\equiv\psi_{\tE_{2s+1}}(x)=
    \dx\!\sum_{0\le j<\frac12(\l-1)}\tilde\rho_{2s+1,j}\,
    \sin(\l-2j-1)\vph(x)\,,
    \label{lame2o}
\end{align}
where now
\begin{equation}
    \tilde\si_{2r,j}=\tp_{2r,j}+\tp_{2r,\l-j-1}\,,\qquad
    \tilde\rho_{2s+1,j}=\tp_{2s+1,j}-\tp_{2s+1,\l-j-1}\,,
    \label{tside}
\end{equation}
and
\begin{equation}
    0\le r\le\left[\frac{\l-1}2\right],\qquad 0\le 
    s\le\left[\frac\l2\right]-1\,.
    \label{rangers2}
\end{equation}

Our next step is to further simplify
Eqs.~\eqref{lame1e}--\eqref{rangers2} for the algebraic
eigenfunctions of the Lam\'e Hamiltonian. To this end, recall that by 
construction the Chebyshev polynomials $T_j$ and $U_j$ ($j=0,1,2,\dots$)
satisfy the identities
\begin{align}
    \cos(j\al)&=T_j(\cos\al)\,,
    \label{tcos}\\
    \sin(j\al)&=\sin\al\,U_{j-1}(\cos\al)\,,
    \label{Uj}
\end{align}
where the latter equality is formally valid for $j=0$ if we set
$U_{-1}=0$. It follows that
\begin{align}
        \ph_{2r}(x)&=
    \sum_{0\le 
    j<\l/2}\si_{2r,j}\,T_{\l-2j}(\cx)
    + p_{2r,\frac\l2}
    \label{phi2r}\\
        \ph_{2s+1}(x)&=
    \sx \sum_{0\le 
    j<\l/2}\rho_{2s+1,j}\,U_{\l-2j-1}(\cx)\,,
    \label{phi2r+1}\\
    \intertext{and, similarly,}
    \tph_{2r}(x)&=\dx\left[\tp_{2r,\frac12(\l-1)}+\sum_{0\le
    j<\frac12(\l-1)}\tilde\si_{2r,j}\,T_{\l-2j-1}(\cx)\right]
    \label{phit2r}\\
    \tph_{2s+1}(x)&= \sx\,\dx \sum_{0\le
    j<\frac12(\l-1)}\tilde\rho_{2s+1,j}\,
    U_{\l-2j-2}(\cx)\,.
    \label{phit2r+1}
\end{align}

It is now straightforward to check that
Eqs.~\eqref{phi2r}--\eqref{phit2r+1} encompass the eight types of
classical Lam\'e polynomials listed in Ref.~\cite{Ar64}. Indeed, note 
first of all that the polynomials $T_j(t)$ and $U_j(t)$ have the parity of
$(-1)^j$ under the reflection $x\mapsto-x$.
Therefore the formulas
\begin{equation}
    \T_{2j}(t)=\frac{T_{2j+1}(t)}{t}\,,
    \quad
    \U_{2j}(t)=\frac{U_{2j+1}(t)}{t}\,;
    \qquad j=0,1,2,\dots\,.
    \label{Ttr}
\end{equation}
define even polynomials $\T_{2j}(t)$ and $\U_{2j}(t)$ of degree
$2j$. If $\l=2N$ ($N=1,2,\dots$), Eqs.~\eqref{phi2r}--\eqref{phit2r+1}
yield the following four types of polynomial --- in $\sn$, $\cn$, 
$\dn$ ---
solutions of the Lam\'e equation:
\begin{align}
    \ph_{2r}(x)&=p_{2r,N}+\sum_{j=0}^{N-1}\si_{2r,j}\,T_{2N-2j}(\cx)\equiv
    \al_r(\sn^2x)
    \label{ph2r2N}\\
    \ph_{2s+1}(x)&=\sx\cx\sum_{j=0}^{N-1}\rho_{2s+1,j}\,\U_{2N-2j-2}(\cx)\equiv
    \sx\cx\,\be_s(\sn^2x)
    \label{ph2s+12N}\\
    \tph_{2s}(x)&=\cx\dx\sum_{j=0}^{N-1}\tilde\si_{2s,j}\,\T_{2N-2j-2}(\cx)\equiv
    \cx\dx\,\tilde\al_s(\sn^2x)
    \label{tph2s2N}\\
    \tph_{2s+1}(x)&=\sx\dx\sum_{j=0}^{N-1}\tilde\rho_{2s+1,j}\,\U_{2N-2j-2}(\cx)\equiv
    \sx\dx\,\tilde\be_s(\sn^2x)\,,
    \label{tph2s+12N}
\end{align}
where
\begin{equation}
    0\le r\le N,\qquad 0\le s\le N-1\,,
    \label{range}
\end{equation}
and $\al_r(t)$, $\be_s(t)$, $\tilde\al_s(t)$, and $\tilde\be_s(t)$ are 
polynomials of respective degrees
\begin{equation}
    \deg\al_r=N\,,
    \quad\deg\be_s=\deg\tilde\al_s=\deg\tilde\be_s=N-1\,.
    \label{degs2N}
\end{equation}
Similarly, if $\l=2N+1$ ($N=0,1,2,\dots$) the exact solutions of the 
Lam\'e equation \eqref{phi2r}--\eqref{phit2r+1} can be written as 
follows:
\begin{align}
    \ph_{2r}(x)&=\cx\sum_{j=0}^{N}\si_{2r,j}\,\T_{2N-2j}(\cx)\equiv
    \cx\,\gamma_r(\sn^2x)
    \label{ph2r2N+1}\\
    \ph_{2r+1}(x)&=\sx\sum_{j=0}^{N}\rho_{2r+1,j}\,U_{2N-2j}(\cx)\equiv
    \sx\,\de_r(\sn^2x)
    \label{ph2r+12N+1}\\
    \tph_{2r}(x)&=\dx\left[\tp_{2r,N}+\sum_{j=0}^{N-1}\tilde\si_{2r,j}\,
    T_{2N-2j}(\cx)\right]\equiv
    \dx\,\tilde\gamma_r(\sn^2x)
    \label{tph2r2N+1}\\
    \tph_{2s+1}(x)&=\sx\cx\dx\sum_{j=0}^{N-1}\tilde\rho_{2s+1,j}\,
    \U_{2N-2j-2}(\cx)\equiv
    \sx\cx\dx\,\tilde\de_s(\sn^2x)\,,
    \label{tph2s+12N+1}
\end{align}
where the range of the indices $r,s$ is still given by
Eq.~\eqref{range}, and $\gamma_r(t)$, $\de_r(t)$,
$\tilde\gamma_r(t)$, and $\tilde\de_s(t)$ are polynomials with
\begin{equation}
    \deg\gamma_r=\deg\de_r=\deg\tilde\gamma_r=N\,,\quad
    \deg\tilde\de_s=N-1\,.
    \label{degs2N+1}
\end{equation}
\subsection{Examples}
\label{subsec.exaint}
The explicit expressions for the Lam\'e polynomials in terms of 
Chebyshev polynomials derived at the end of the previous section can be used to 
algorithmically compute the Lam\'e polynomials and their corresponding energies for 
low values of $\l$.

When $\l=1$, Eqs.~\eqref{ph2r2N+1}--\eqref{tph2s+12N+1} for $N=0$ immediately
yield the following well-known formulas for the Lam\'e
polynomials of order $1$:
\begin{equation}
    \ph_0(x)\propto\cx,\qquad\ph_1(x)\propto\sx,\qquad
    \tph_0\propto\dx\,.
    \label{l=1}
\end{equation}
The critical polynomials $\hat P_2$ and $\htP_1$ are easily computed 
using the recurrence relations \eqref{rr}--\eqref{tbj}, with the result
\begin{equation}
    \hat P_2(E)=E^2-(m+2)E+m+1\,,\qquad
    \htP_1(E)=E-m\,.
    \label{pcrit1}
\end{equation}
The energies associated to the eigenfunctions \eqref{l=1} are just 
the roots of each of the latter polynomials ordered increasingly, namely
\begin{equation}
    E_0=1\,,\qquad E_1=m+1\,;\qquad \tE_0=m\,.
    \label{El=1}
\end{equation}

For $\l=2$ the critical polynomials are
\begin{align}
    \hat P_3(E)&=E^3 -(5 m + 8) E^2 + 4(m^2 + 8m + 4) E - 12 m\,(m + 
    4)\,,\label{pcrit2}\\
\htP_2(E)&=E^2 -(5 m + 2) E + (m + 1) (4m + 1)\,.
    \label{pcrit2t}
\end{align}
The energies corresponding to the Lam\'e polynomials of order two are 
therefore given by
\begin{gather}
    \label{El=2}
    \left.
    \begin{aligned}
    E_0& \\
    E_2&
    \end{aligned}
     \right\}
    =2 \left(1 + m \mp \sqrt{m^2-m+1} \right)\,,\qquad
    E_1=4+ m\,;\\
   \tE_0 = m+1,\qquad \tE_1 = 4\,m+1\,.
    \label{Etl=2}
\end{gather}
Note that the factor of $2$ in \eqref{El=2} is missing in 
Ref.~\cite{Ar64}, p.~205. The Lam\'e polynomials are easily computed using 
\eqref{ph2r2N}--\eqref{tph2s+12N} with $N=1$, with the following result:
\begin{gather}
    \left.
    \begin{aligned}
	-\frac{\ph_0}4&\\
	-\frac{\ph_2}4&
    \end{aligned}
    \right\}
	= \sn^2 x-\frac1{3m}\left(1 + m \pm \sqrt{m^2-m+1}\right),
    \label{ph02l2}\\
     \frac{\ph_1}4=\sx\,\cx\,;\\
     \frac{\tph_0}2=\cx\,\dx\,,\qquad
     \frac{\tph_1}4=\sx\,\dx\,.
    \label{phi1l2}
\end{gather}

When $\l=3$ the critical polynomials are given by
\begin{align}
    \hat P_4(E)&=\left[E^2 - 2 ( 5 + 2\,m ) E + 3 ( 3 + 
    8\,m)\right]\notag\\
    &\hspace{1cm}{}\times
    \left[E^2-10 ( 1 + m) E+ 3(3\,m^2+26\,m +3) \right],
    \label{pcrit3}\\
    \htP_3(E)&=\left[E-4(m+1)\right]\left[E^2-2(2+5m)E+3m(3m+8\right],
    \label{pcrit3t}
\end{align}
from which the following exact energies are computed:
\begin{align}
        \left.
    \begin{gathered}
    E_0\\
    E_2
    \end{gathered}
     \right\}
    &=2m+5\mp2\sqrt{m^2-m+4}\,,
\label{E02l3}
    \\
    \left.
        \begin{gathered}
    E_1 \\
    E_3
    \end{gathered}
     \right\}
    &=5(m+1)\mp2\sqrt{4m^2-7m+4}\,;
    \label{E13l3}\\
    \left.
    \begin{gathered}
    \tE_0 \\
    \tE_2
    \end{gathered}
     \right\}
    &=5m+2\mp2\sqrt{4m^2-m+1}\,,
    \label{tE02l3}\\
    \tE_1&=4(m+1)\,.
    \label{tE1l3}
\end{align}
From Eqs.~\eqref{ph2r2N+1}--\eqref{tph2s+12N+1} with $N=1$, the
corresponding Lam\'e polynomials are given by
\begin{align}
    \left.
    \begin{gathered}
	-\frac{\ph_0}8\\
	-\frac{\ph_2}8
    \end{gathered}
    \right\}
    &=\cx\left[\sn^2x-\frac1{5m}\left(m+2\pm\sqrt{m^2-m+4}\right)\right],
    \label{ph02l3}\\
    \left.
    \begin{gathered}
        -\frac{\ph_1}8\\
	-\frac{\ph_3}8 
    \end{gathered}
    \right\}
    &=\sx\left[\sn^2x-\frac1{5m}\left(2(m+1)\pm\sqrt{4m^2-7m+4}\right)\right];
    \label{ph13l3}\\
    \left.
    \begin{gathered}
	-\frac{\tph_0}4\\
	-\frac{\tph_2}4
    \end{gathered}
    \right\}
    &=\dx\left[\sn^2x-\frac1{5m}\left(2m+1\pm\sqrt{4m^2-m+1}\right)\right],
    \label{tph02l3}\\
    \frac{\tph_1}4&=\sx\,\cx\,\dx\,.
    \label{tph1l3}
\end{align}
The latter formulas for the Lam\'e polynomials of order three and 
their corresponding energies are seen to coincide with those given in 
Ref.~\cite{Ar64}.

When $\l$ is greater than three, it becomes increasingly more 
difficult to compute the Lam\'e polynomials of order $\l$ and their 
corresponding eigenfunctions in closed form. This is essentially due 
to the fact that the high degree of the critical polynomials $\hat P_\l$ and 
$\htP_{\l+1}$ in general makes it impossible to exactly compute the energies $E_i$ 
and $\tE_i$, which enter in the definition of the coefficients 
$p_{ij}$ and $\tilde p_{ij}$.
For example, for $\l=4$ the critical polynomials $\hat P_5$ and $\htP_4$ 
are given by
\begin{align}
    \hat P_5(E)&=\left[E^2- 10\,E\,(2 + m)+ 64 + 136\,m + 9\,m^2 \right]\nt
   &{}\times\left[ E^3 - 20\,E^2\,(1 + m)
    + 16\,E\,(4 + 21\,m +
    4\,m^2)- 640\,m\,(1 + m)\right],
    \label{pcrit4}\\
    \htP_4(E)&=\left[E^2 - 10\,E\,(1 + 
    m) + 9  + 46\,m + 9\,m^2\right]\nt
    &\qquad{}\times\left[E^2- 10\,E\,(1 + 2\,m) + 9 + 136\,m + 64\,m^2 \right].
    \label{tpcrit4}
\end{align}
It is not difficult to verify that the roots of the quadratic factor of
$P_5$ are the energies $E_1$ and $E_3$, so that we have the exact 
formula
\begin{equation}
    \left.
    \begin{gathered}
    E_1\\
    E_3
    \end{gathered}
    \right\}
    = 5(m+2)\mp\sqrt{4m^2-9m+9}\,.
    \label{E13l4}
\end{equation}
The corresponding Lam\'e polynomials are easily computed using 
Eq.~\eqref{ph2s+12N}:
\begin{equation}
    \left.
    \begin{gathered}
    -\frac{\ph_1}{16}\\
    -\frac{\ph_3}{16}
\end{gathered}
    \right\}
    =\sx\,\cx \left[
    \sn^2x-\frac1{7m}\left(2m+3\pm\sqrt{4m^2-9m+9}\right)
    \right].
    \label{ph13l4}
\end{equation}
Similarly, the fact that $\tP_4$ factorizes into two quadratic 
polynomials allows us to exactly compute the energies $\tE_i$ in this 
case, obtaining
\begin{align}
        \left.
    \begin{gathered}
    \tE_0\\
    \tE_2
    \end{gathered}
    \right\}
    &=5(m+1)\mp2\sqrt{4m^2+m+4}\,,
    \label{tE02l4}\\
            \left.
    \begin{gathered}
    \tE_1\\
    \tE_3
    \end{gathered}
    \right\}
    &=5(2m+1)\mp2\sqrt{9m^2-9m+4}\,.
    \label{tE13l4}
\end{align}
By Eqs.~\eqref{tph2s2N}--\eqref{tph2s+12N}, the corresponding
eigenfunctions are given by
\begin{align}
        \left.
    \begin{gathered}
    -\frac{\tph_0}8\\
    -\frac{\tph_2}8
    \end{gathered}
    \right\}
    &= \cx\,\dx \left[
    \sn^2x-\frac1{7 m}\left(2(m+1)\pm\sqrt{4m^2+m+4}\right)
    \right],
    \label{tph02l4}\\
            \left.
    \begin{gathered}
    -\frac{\tph_1}{16}\\
    -\frac{\tph_3}{16}
    \end{gathered}
    \right\}
    &= \sx\,\dx \left[
    \sn^2x-\frac1{28\,m}\left(5m+8\pm4\sqrt{9m^2-9m+4}\right)
    \right].
    \label{tph3l4}
\end{align}
The remaining Lam\'e polynomials of order four can be expressed using
Eq.~\eqref{ph2r2N} in terms of the three roots $E_i$ ($i=0,2,4$) of
the cubic factor of $\hat P_5$, with the following result:
\begin{multline}
    \frac{\ph_i}{16}=\sn^4x-\frac1{14\,m}\left[16(m+1)-E_i\right]\,\sn^2x\\
    {}+
    \frac1{280\,m^2}\left[E_i^2-20(m+1)E_i+64m^2+296m+64\right].
    \label{ph024l4}
\end{multline}
For instance, for $m=1/2$ the roots $E_0$, $E_2$ and $E_4$ can be 
computed in closed form:
\begin{equation}
            \left.
    \begin{gathered}
    E_0\\
    E_4
    \end{gathered}
    \right\}
    = 10\mp2\sqrt{13}\,,
   \qquad
    E_2=10\,.
    \label{E2l4}
\end{equation}
Substituting into Eq.~\eqref{ph024l4} we obtain the following exact 
expressions 
for the corresponding Lam\'e polynomials for $m=1/2$:
\begin{gather}
    \left.
    \begin{aligned}
    \frac{\ph_0}{16}\\
    \frac{\ph_2}{16}
    \end{aligned}
    \right\}
    = \sn^4x-\frac27\left(7\pm\sqrt{13}\right)\,\sn^2x
    +\frac27\left(4\pm\sqrt{13}\right)\,,
    \label{ph04l4m12}\\
    \frac{\ph_2}{16}=\sn^4x-2\,\sn^2x+\frac25\,.
    \label{ph2l4m12}
\end{gather}
\section{Case II: $\l$ is a positive half-integer}
\label{subsec.half}
This is the most interesting case, since we shall see that it leads 
to the non-meromorphic Lam\'e functions, whose algebraization has 
proved more difficult than that of the Lam\'e polynomials.

The gauge factor can be computed from Eqs.~\eqref{geq},
\eqref{zrule}, and \eqref{alg3}--\eqref{alg4}, with the result
\begin{equation}
        \mu\left(\z(x)\right)=\cn^n x\,(\cx\pm\ii\,k'\sx)^{1/2}\,.
    \label{gl2}
\end{equation}
The upper and lower signs in this formula correspond, respectively, 
to the solutions \eqref{alg3} and \eqref{alg4},
\begin{equation}
    n=\l-\frac12\,,
    \label{n}
\end{equation}
and we are using the customary notation
\begin{equation}
    k=\sqrt m,\qquad k'=\sqrt{1-m}\,.
    \label{k'}
\end{equation}
Thus when $\l$ is a positive half-integer the Lam\'e equation has 
$2n+2=2\l+1$ algebraic solutions of the form
\begin{equation}
    \psi^\pm(x) = \cn^n x\,(\cx\pm\ii\,k'\sx)^{1/2}\,\chi^\pm\left(\frac{\sn 
    x}{\cx}\right)\,,
    \label{sol34}
\end{equation}
where $\chi^\pm$ is a polynomial of degree at most $n$.

We shall now use the same strategy as in the previous section to simplify 
the eigenfunctions \eqref{sol34}.
It is straightforward to check that in this case only the two roots
$\pm\ii/k'$ of the polynomial $P(z)$ in \eqref{P} satisfy \eqref{cond}. For reasons
that shall become clear later, we shall choose
$$
z_1=\mp\frac\ii{k'},\quad z_2=\pm\frac\ii{k'}\,,
$$
where again the upper sign corresponds to \eqref{alg3} and the lower
sign to \eqref{alg4}. Using the same convention for the eigenfunctions
$\psi_{E^\pm_i}^\pm$ and the polynomials $P_j^\pm$, a straightforward
calculation yields
\begin{equation}
    \psi_{E^\pm_i}^\pm(x)=\sum_{j=0}^n \frac{(-1)^j}{j!}P_j^\pm(E^\pm_i)\,
    (\cx\pm\ii\,k'\sx)^{n-j+\frac12}(\cx\mp\ii\,k'\sx)^j\,.
    \label{eigenf}
\end{equation}
We have dropped an inessential constant factor in the latter formula, 
and the energies $E^\pm_i$ are defined by
\begin{equation}
  P^\pm_{n+1}(E^\pm_i)=0,\qquad 0\le i\le n\,.
    \label{epm}
\end{equation}
(We shall prove below that the roots of $P^\pm_{n+1}$ are indeed
real and simple.)

We now compute the polynomials $P^\pm_j$, or more precisely the 
recurrence relation defining them. In principle, we should use the notation $\hat P^\pm(w)$ and $\hat Q^\pm(w)$ 
to denote the polynomials $\hat P(w)$, $\hat Q(w)$ in 
Eqs.~\eqref{phat}--\eqref{qhat} corresponding to each of the 
two cases \eqref{alg3} and \eqref{alg4}. Remarkably, a direct 
calculation shows that this is actually unnecessary, since
\begin{align}
    \hat P^+(w)=\hat P^-(w)\equiv \hat P(w) &= 
    -\left[m\,w^3+2\,(2-m)\,w^2+m\,w\right].
    \label{phat2}\\
    \hat Q^+(w)=\hat Q^-(w)\equiv\hat Q(w) &= \frac 
    m2\,(n+2)\,w^2+(2-m)\,w-\frac12 m n\,.
    \label{qhat2}
\end{align}
This implies that the recurrence relations determining the two sets of
orthogonal polynomials $\left\{P^\pm_j(E)\right\}_{j\ge0}$ are in fact
the same\footnote{From Eq.~\eqref{c*} we have $c_*^+=c_*^-$, and from
the fact that $R$ is invariant under projective transformations,
\cite{GKOnorm93}, and Eqs.~\eqref{PQR} and \eqref{phat2} we obtain
$$
\hat c_*^+=\frac n{12}(n+2)\left(c_{00}-\hat c_{00}^+\right) + c_*^+=\hat c_*^-
\,.
$$}.
Since we have normalized both sets so that
$P^\pm_0(E)=1$, it follows that both polynomial families are
identical, namely
\begin{equation}
    P^+_j(E)=P^-_j(E)\equiv P_j(E),\qquad j=0,1,2,\dots\,.
    \label{Ppmeq}
\end{equation}
In particular, the above equality for $j=n+1$ and Eq.~\eqref{epm} 
show that
\begin{equation}
    E^+_i=E^-_i\equiv E_i,\qquad 0\le i\le n\,.
    \label{Epmeq}
\end{equation}
Moreover, to each of these $n+1$ energies correspond two 
algebraic eigenfunctions of the form \eqref{eigenf},
which on account of \eqref{Ppmeq} can be written as
\begin{equation}
    \psi^\pm_{E_i}=\sum_{j=0}^n \frac{(-1)^j}{j!}P_j(E_i)\,
    (\cx\pm\ii\,k'\sx)^{n-j+\frac12}(\cx\mp\ii\,k'\sx)^j\,.
    \label{psipmei}
\end{equation}
From Eqs.~\eqref{phat2}--\eqref{qhat2} we see that the coefficients of
the polynomials $\hat P(w)$ and $\hat Q(w)$ are all real. This
implies, \cite{FGRorth96}, that the coefficients of the recurrence
relation determining the polynomials $P_j(E)$ are also real. Since
$P_0(E)=1$, it follows that $P_j(E)$ is a polynomial with real
coefficients, for all $j=0,1,\dots$. From Eq.~\eqref{psipmei} and the
positivity of the coefficient $a_j$ for $1\le j\le n$ 
(cf.~Eq.~\eqref{aj} below), it follows that
all the energies $E_i$ ($0\le i\le n$) are real and simple, and therefore
$$
\psi^-_{E_i}=\overline{\psi^+_{E_i}},\qquad 0\le i\le n\,,
$$
where the overbar denotes complex conjugation. If we set
$$
\psi_i(x)\equiv\psi^+_{E_i}(x)\,,\qquad 0\le i\le n\,,
$$
then for each root $E_i$ of $P_{n+1}$ the Lam\'e equation has \emph{two}
algebraic \emph{real}-valued solutions given by
\begin{equation}
    \phi^1_i(x)=\Re\psi_i(x)\,,\qquad \phi^2_i(x)=\Im\psi_i(x)\,.
    \label{realsol}
\end{equation}
The recurrence relation satisfied by the polynomials $P_j(E)$ can be 
easily written down from \eqref{phat2}--\eqref{qhat2} using the 
prescription of Ref.~\cite{FGRorth96}. Indeed, if, following the 
latter reference, we write
\begin{equation}
    \frac{P_j(E)}{j!}=\left(\frac 2k\right)^{2j}\frac{\hat 
    P_j(E)}{(2j)!}\,,
    \label{pphat}
\end{equation}
then the polynomials $\hat P_j$ satisfy a three-term recurrence relation
of the form \eqref{rr}, with
\begin{align}
    a_j&=\frac 14\,m^2 j(2j-1)(2n-2j+3)(n-j+1)\,,
    \label{aj}\\
    b_j&=\frac14\,(2n+1)(m+2n+1)-(2-m)j\,(2n-2j+1)\,.
    \label{bj}
\end{align}

We are ready to show that, when $\l$ is a half-integer, the
$2\l+1$ algebraic solutions of the Lam\'e equation of
the form \eqref{realsol}--\eqref{psipmei} obtained in the previous section are
precisely the non-meromorphic Lam\'e functions.

To this end, we need to simplify Eq.~\eqref{psipmei}. In the first 
place, since
\begin{equation}
    \abs{\cx\pm\ii\,k'\sx}=\dx
    \label{mod}
\end{equation}
(recall that $\dx$ is positive for all real $x$), we can write
\begin{equation}
    \cx\pm\ii\,k'\sx=\dx\,\e^{\pm\ii\,\th(x)}\,,
    \label{cnsn}
\end{equation}
so that
\begin{equation}
    \cos\th(x)=\frac{\cx}{\dx}\,,\qquad
    \sin\th(x)=k'\frac{\sx}{\dx}\,.
    \label{theta}
\end{equation}
Secondly, we have
\begin{equation}
    (\cx+\ii\,k'\sx)^{1/2}=\pm\frac1{\sqrt2}\,\left(\sqrt{\dx+\cn
    x}+\ii\,\ep(x)\sqrt{\dx-\cx}\right),
    \label{sqrt}
\end{equation}
where we have set
\begin{equation}
    \ep(x)=\sign(\sx)\,.
    \label{eps}
\end{equation}
Note that the functions $\sqrt{\dx+\cx}$ and $\ep(x)\,\sqrt{\dn
x-\cx}$ are $C^\infty$ on the whole real line except at the points
$x=2(2r+1)\,K$. We shall also find useful in the sequel the 
identities
\begin{equation}
    \sqrt{\dx\pm\cx}=\ep(x)\,\frac{\dx\pm\cx}{k'\sx}\,\sqrt{\dx\mp\cx}\,.
    \label{sqrtid}
\end{equation}

From \eqref{cnsn} it follows that
\begin{multline}
    \sum_{j=0}^n p_{ij}\,(\cx+\ii\,k'\sx)^{n-j}(\cx-\ii\,k'\sn 
    x)^j\\{}=
    \dn^n x\,\sum_{j=0}^n p_{ij}\,\e^{\ii\,(n-2j)\th(x)}\,,
    \label{sumsimp}
\end{multline}
where the coefficients $p_{ij}$ are again defined by Eq.~\eqref{pptij}.
The sum in the latter equation is of the same time as \eqref{psiphi}, 
and therefore can be expressed as follows:
\begin{align}
    \sum_{j=0}^n p_{ij}\,\e^{\ii\,(n-2j)\th(x)}
    &= p_{i,n/2}+\sum_{0\le j<n/2}
    \si_{ij}\,T_{n-2j}\bigl(\cos\th(x)\bigr)\nt
    &\qquad{}+
    \ii\,\sin\th(x) \sum_{0\le j<n/2}
    \rho_{ij}\,U_{n-2j-1}\bigl(\cos\th(x)\bigr)\nt
    &=p_{i,n/2}+\sum_{0\le j<n/2}
    \si_{ij}\,T_{n-2j}(\cd)\nt
    &\qquad{}+
    \ii\,k'\,\sd \sum_{0\le j<n/2}
    \rho_{ij}\,U_{n-2j-1}(\cd)
    \label{sumodd}
\end{align}
where we have used the customary abbreviations
$$
\cd\equiv \frac{\cx}{\dx}\,,\qquad
\sd\equiv \frac{\sx}{\dx}\,,
$$
and we have again set
\begin{equation}
    \si_{ij}=p_{ij}+p_{i,n-j}\,,\qquad
    \rho_{ij}=p_{ij}-p_{i,n-j}\,.
    \label{sidehalf}
\end{equation}
From Eqs.~\eqref{sqrt} and \eqref{sqrtid} (dropping the constant
factor $\pm1/\sqrt2$) we easily obtain
\begin{multline}
    \phi^1_i(x)=
    \sqrt{\dx+\cx}\,\dnx\left[p_{i,n/2}+\sum_{0\le j<n/2}
    \si_{ij}\,T_{n-2j}(\cd)\right.\\
    \left.\quad{}+(\cd-1)
    \sum_{0\le j<n/2}\rho_{ij}\,U_{n-2j-1}(\cd)
    \right].
    \label{ph1}
\end{multline}
An analogous
calculation leads to the following expression for the second solution
$\phi^2_i(x)$:
\begin{multline}
    \phi^2_i(x)=
    \ep(x)\,\sqrt{\dx-\cx}\,\dnx\left[p_{i,n/2}+\sum_{0\le j<n/2}
    \si_{ij}\,T_{n-2j}(\cd)\right.\\
    \left.\quad{}+(\cd+1)
    \sum_{0\le j<n/2}\rho_{ij}\,U_{n-2j-1}(\cd)
    \right].
    \label{ph2}
\end{multline}
    
Let us now study in more detail the properties of the $2\l+1$
solutions of the Lam\'e equation of the form
\eqref{ph1}--\eqref{ph2} that we have just obtained. From the
differentiability properties of the functions $\sqrt{\dx+\cx}$ and
$\ep(x)\sqrt{\dx-\cx}$ mentioned above, it follows that $\phi^1_i(x)$
and $\phi^2_i(x)$ are $C^\infty$ solutions of the Lam\'e equation
\eqref{lame} with $E=E_i$ in the open interval $(-2K,2K)$, and are
clearly continuous on $[-2K,2K]$.
%
%
The product of $\dnx$ with
the functions within square brackets in Eqs.~\eqref{ph1} and
\eqref{ph2} defines two $C^\infty$ functions on the whole
real line (they are in fact polynomials in $\sx$, $\cx$, $\dx$), with
(real) period $4K$. On the other hand, it is immediate to check that
if we prolong $f(x)=\sqrt{\dx+\cx}$ or $f(x)=\ep(x)\sqrt{\dx-\cx}$
outside the interval $[-2K,2K]$ anti-periodically, i.e. requiring that
\begin{equation}
    f(x+4K)=-f(x)\,,
    \label{antiper}
\end{equation}
then $f$ becomes a $C^\infty$ function. (Indeed, in a neighborhood of
$\pm 2K$ the prolongations of $\sqrt{\dx+\cx}$ and
$\ep(x)\sqrt{\dx-\cx}$ are equal, respectively, to $\pm
k'\sx/\sqrt{\dx-\cx}$ and $\pm\sqrt{\dx-\cx}$, the latter functions
being obviously $C^\infty$ at $\pm2K$.) It follows that, if we define
$\phi^1_i(x)$ and $\phi^2_i(x)$ outside $[-2K,2K]$ by the prescription
\eqref{antiper}, then $\phi^1_i$ and $\phi^2_i$ are $C^\infty$
on the whole real line, and have (real) period $8\,K$. Furthermore,
from Eqs.~\eqref{ph1} and \eqref{ph2} it follows that
$\ph_i^1(x)$ and $\ph_i^2(x)$ are, respectively, even and odd
functions of the variable $x$. It is also clear from the 
well-known equalities
$$
\sn(x+2K)=-\sx\,,\quad \cn(x+2K)=-\cx\,,\quad \dn(x+2K)=\dx\,,
$$
that $\ph^1_i$ and $\ph^2_i$ are related by
\begin{equation}
    \ph^2_i(x)=\ph^1_i(x+2K)\,.
    \label{ph12odd}
\end{equation}
%
%
%
%
%
From this equality and the anti-periodicity condition \eqref{antiper} 
it follows that
\begin{equation}
    \bigl[\ph^1_i(x+2K)\pm\ii\,\ph^2_i(x+2K)\bigr]=\mp\ii\,
    \bigl[\ph^1_i(x)\pm\ii\,\ph^2_i(x)\bigr]\,,
    \label{qm}
\end{equation}
so that the quasi-momentum associated to each of the $2\l+1$ 
algebraic energies 
$E_i$ is equal to $\pi/(4K)$. 

When $n$ is odd, it is convenient to rewrite Eq.~\eqref{ph1} in 
the following way:
    \begin{align}
    \phi^1_i(x)&=
    \sqrt{\dx+\cx}\,\dnx\left[\cd\,\sum_{j=0}^{\frac12(n-1)}
    \si_{ij}\,\T_{n-2j-1}(\cd)\right.\nt
    &\left.\kern8em{}+(\cd-1)
    \sum_{j=0}^{\frac12(n-1)}\rho_{ij}\,U_{n-2j-1}(\cd)
    \right]\nt
    &=\sqrt{\dx+\cx}\,\dn^{n-1}x\left[\cx\,\sum_{j=0}^{\frac12(n-1)}
    \si_{ij}\,\T_{n-2j-1}(\cd)\right.\nt
    &\left.\kern8em{}+(\cx-\dx)
    \sum_{j=0}^{\frac12(n-1)}\rho_{ij}\,U_{n-2j-1}(\cd)
    \right].
    \label{ph1odd}
\end{align}
Taking into account the parity properties of the Chebyshev 
polynomials, and the fact that $\cn^2x$ and $\dn^2x$ can be expressed 
linearly in terms of $\sn^2x$, it follows that
\begin{equation}
    \ph^1_i(x)=\sqrt{\dx+\cx}\left[\cx\,\al_i(\sn^2x)
    +\dx\,\be_i(\sn^2x)\right],
    \label{phi1ince}
\end{equation}
where $\al_i(t)$ and $\be_i(t)$ are polynomials in $t$ of degree
$\frac12(n-1)$ given by
\begin{align}
    \al_i(\sn^2x)&=\dn^{n-1} x\sum_{j=0}^{\frac12(n-1)}
    \left\{
    \si_{ij}\,\T_{n-2j-1}(\cd)
    +\rho_{ij}\,U_{n-2j-1}(\cd)
    \right\},
    \label{ai}\\
    \be_i(\sn^2x)&=
    -\dn^{n-1} x
    \sum_{j=0}^{\frac12(n-1)}
    \rho_{ij}\,U_{n-2j-1}(\cd)\,.
    \label{bi}
\end{align}
Comparing \eqref{phi1ince} with Ref.~\cite{In39} (p.~97),
we see that $\ph^1_i$ is proportional to the non-meromorphic Lam\'e 
function $\Ec_\l^{r_i+\frac12}$,
\begin{equation}
    \ph^1_i(x)\propto \Ec_\l^{r_i+\frac12}\,,\qquad 0\le i\le 
    n\equiv\l-\frac12\,,
    \label{ecl}
\end{equation}
where $r_i$ is the number of zeros of $\ph^1_i$ in the interval
$(0,2K)$. From Eq.~\eqref{ph2} or, alternatively, from
Eqs.~\eqref{ph12odd}) and \eqref{phi1ince}, we also obtain
\begin{equation}
    \ph^2_i(x)=\ep(x)\sqrt{\dx-\cx}\left[\cx\,\al_i(\sn^2x)
    -\dx\,\be_i(\sn^2x)\right],
    \label{phi2ince}
\end{equation}
so that
\begin{equation}
    \ph^2_i(x)\propto \Es_\l^{r_i+\frac12}\,,\qquad 0\le i\le
    n\equiv\l-\frac12\,.
    \label{esl}
\end{equation}
%
%

Similarly, when $n$ is even we rewrite Eq.~\eqref{ph1}
as follows:
\begin{align}
    \phi^1_i(x)&=
    \sqrt{\dx+\cx}\,\dnx\left[p_{i,n/2}+\sum_{j=0}^{\frac n2-1}
    \si_{ij}\,T_{n-2j}(\cd)\right.\nt
    &\left.\kern8em{}+\cd\,(\cd-1)
    \sum_{j=0}^{\frac n2-1}\rho_{ij}\,\U_{n-2j-2}(\cd)
    \right]\nt
    &=\sqrt{\dx+\cx}\,\dn^{n-2}x\left[\dn^2x\Bigl(p_{i,n/2}+
    \sum_{j=0}^{\frac n2-1}
    \si_{ij}\,T_{n-2j}(\cd)\Bigr)\right.\nt
    &\left.\kern8em{}+\cx\,(\cx-\dx)
    \sum_{j=0}^{\frac n2-1}\rho_{ij}\,\U_{n-2j-2}(\cd)
    \right].
    \label{ph1even}
\end{align}
We thus obtain
\begin{equation}
    \ph^1_i(x)=\sqrt{\dx+\cx}\,\left[\al_i(\sn^2
    x)+\cx\,\dx\,\be_i(\sn^2 x)\right],
    \label{ph1eince}
\end{equation}
where $\al_i(t)$ and $\be_i(t)$ are polynomials in $t$ of respective 
degrees $\frac n2$ and $\frac n2-1$, defined by
\begin{align}
    \al_i(\sn^2x)&=\dnx\left[p_{i,n/2}+\sum_{j=0}^{\frac n2-1}
    \si_{ij}\,T_{n-2j}(\cd)\right]\nt
    &\quad{}+\dn^{n-2}x\,\cn^2x\,\sum_{j=0}^{\frac 
    n2-1}\rho_{ij}\,\U_{n-2j-2}(\cd)\,,
    \label{aie}\\
    \be_i(\sn^2x)&=
    -\dn^{n-2} x
    \sum_{j=0}^{\frac12(n-1)}
    \rho_{ij}\,\U_{n-2j-2}(\cd)\,.
    \label{bie}
\end{align}
From \eqref{ph12odd} we have
\begin{equation}
    \ph^2_i(x)=\ep(x)\,\sqrt{\dx-\cx}\,\left[\al_i(\sn^2
    x)-\cx\,\dx\,\be_i(\sn^2 x)\right].
    \label{ph2eince}
\end{equation}
Comparing, again, with Ref.~\cite{In39} we deduce that 
Eqs.~\eqref{ecl} and \eqref{esl} also hold when $n$ is an even 
non-negative integer.
\subsection{Examples}
\label{subsec.exahalf}
The case $n=0$, i.e, $\l=1/2$, is particularly easy to deal with in the 
framework of our formalism. Indeed, in this case we can exactly 
compute one eigenvalue $E_0$ of the Lam\'e Hamiltonian, its 
corresponding two linearly independent eigenfunctions given by 
\eqref{ph1eince}--\eqref{ph2eince} with $n=0$:
\begin{equation}
    \ph^1_0(x)=\sqrt{\dx+\cx}\,,\qquad 
    \ph^2_0(x)=\ep(x)\sqrt{\dx-\cx}\,.
    \label{zeroef}
\end{equation}
The eigenvalue $E_0$ is the root of the first degree polynomial $P_1$ 
or, equivalently, of $\hat P_1$ (since both polynomials differ by a 
constant factor, by \eqref{pphat}). From Eq.~\eqref{rr} with $j=0$ we 
get
$$
\hat P_1(E)=(E-b_0)\,\hat P_0=E-b_0\,,
$$
and from \eqref{bj} (with $n=0$) it follows that
\begin{equation}
    E_0=b_0=\frac14\,(1+m)\equiv\frac14\,(1+k^2)\,.
    \label{E0}
\end{equation}
We have thus shown in a purely algebraic fashion that the general
solution of the Lam\'e equation
$$
\psi''(x)+\frac14\,\left(1+k^2-3\,k^2 \sn^2 x\right)\,\psi(x)=0
$$
is a linear combination of the two functions \eqref{zeroef}.

Similarly, for $n=1$, i.e, $\l=3/2$, the two algebraic
eigenvalues of the Lam\'e Hamiltonian are the roots of the polynomial
\begin{equation}
    \hat P_2(E)=E^2-\frac52\,(m+1)\,E+\frac3{16}\,(3\,m^2+23\,m+3)\,,
    \label{crit1}
\end{equation}
namely,
\begin{equation}
    \left.
    \begin{split}
    &E_0\\
    &E_1
    \end{split}
    \right\}
    =\frac54\,(m+1)\mp\sqrt{m^2-m+1}\,.
    \label{eig1}
\end{equation}
Their associated eigenfunctions are easily computed using
Eqs.~\eqref{phi1ince}--\eqref{bi}, which for $n=1$ reduce to
\begin{align*}
\al_i&=2\,p_{i0}=2\,,\\
\be_i&=p_{i1}-p_{i0}=-P_1(E_i)-1=-\frac2m\,\hat P_1(E_i)-1
= \frac2m\,\left[\frac34\,(m+3)-E_i\right]-1\,.
\end{align*}
From Eq.~\eqref{phi1ince} we obtain
\begin{equation}    
    \left.
    \begin{split}
	&\ph_0^1\\
	&\ph_1^1
    \end{split}
    \right\}
    =\sqrt{\dx+\cx}\,
    \left[m\,\cx +\left(1-m\pm\sqrt{m^2-m+1}\right)\dx\right],
    \label{eigfn1}
\end{equation}
where  we have
omitted a trivial constant factor. The remaining two eigenfunctions
$\ph^2_0$ and $\ph^2_1$ can be easily obtained from \eqref{eigfn1}
using \eqref{phi2ince}:
\begin{equation}    
    \left.
    \begin{split}
	&\ph_0^2\\
	&\ph_1^2
    \end{split}
    \right\}
    =\ep(x)\,\sqrt{\dx-\cx}\,
    \left[m\,\cx -\left(1-m\pm\sqrt{m^2-m+1}\right)\dx\right].
    \label{eigfn2}
\end{equation}
The latter formulas are easily seen to coincide with the ones at the 
end of Ref.~\cite{In39}.

As in the case of integer $\l$, the computation of the non-meromorphic
Lam\'e functions in closed form quickly becomes unmanageable as the
degree of the critical polynomial $\hat P_{n+1}$ increases. For
example, for $\l=5/2$ (i.e., $n=2$) the algebraic eigenfunctions of
the form \eqref{ph1even} can be expressed as follows
\begin{equation}
    \begin{split}
    \frac{\ph_i^1}{24\,m^2}&=\sqrt{\dx+\cx}\left\{
    -25 - 130\,m + 87\,m^2 + ( 104 + 40\,m) \,E_i-16\,E_i^2\right.\\
     &{}+m\,\sn^2x\,\left[ 25 + 34\,m + 9\,m^2 - 
     (104+40\,m) \,E_i + 
     16\,E_i^2 \right]
    \\
    &{}+\left. 2\, \cx\,\dx\,\left[ 25 + 430\,m + 21\,m^2  - 
     8\,( 13 + 11\,m ) \,E_i + 
     16\,E_i^2 \right]
    \right\},
    \end{split}
    \label{phi5/2}
\end{equation}
where $E_i$ is one of the three real roots of the critical polynomial
\begin{multline}
    \hat P_3(E)=E^3-\frac{35}4(m+1)E^2+\frac7{16}\left(
    37m^2+138m+37\right)\!E\\
    {}-\frac5{64}(m+1)\left(45m^2+794m+45\right).
    \label{pcritl5/2}
\end{multline}
The remaining three eigenfunctions $\ph_i^2$ ($i=0,1,2$) are obtained 
from the Eq.~\eqref{phi5/2} using Eq.~\eqref{ph2eince}. When $m=1/2$, the 
energies $E_i$ can again be computed explicitly, i.e.,
\begin{equation}
    \left.
    \begin{gathered}
	E_0\\
	E_2
    \end{gathered}
    \right\}
    = \frac{35}8\mp\sqrt 7\,,\qquad
    E_1=\frac{35}8\,.
    \label{El5/2}
\end{equation}
Consequently, Eq.~\eqref{phi5/2} for the eigenfunctions yields the 
following explicit expressions:
\begin{gather}
        \left.
    \begin{gathered}
	-\frac34\,\ph_0^1\\
	-\frac34\,\ph_2^1
    \end{gathered}
    \right\}
     = \sqrt{\dx+\cx}\left[(5\pm\sqrt7)\,\sn^2x-2\,(2\pm\sqrt7)\,\cx\,\dx
     -7\mp2\sqrt7\right],
    \label{phi02l5/2m1/2}\\
    -\frac1{16}\ph_1^1=\sqrt{\dx+\cx}\left(\sn^2 x+2\,\cx\,\dx-\frac74\right).
\end{gather}


\frenchspacing
\begin{thebibliography}{99} 
\bibitem{In39a}
Ince E L 1939--40 {\em Proc. Roy. Soc. Edinb.} {\bf 60} 47--63
\bibitem{In39}
Ince E L 1939--40 {\em Proc. Roy. Soc. Edinb.} {\bf 60} 83--99
\bibitem{EMOT55}
Erd{\'e}lyi A, Magnus W, Oberhettinger F and Tricomi F G 1955 {\em Higher
Transcendental Functions}, vol.~III (New York: Mc-Graw Hill)
\bibitem{Ar64}
Arscott F M 1964 {\em Periodic Differential Equations,} (Oxford: Pergamon)
\bibitem{AGI83}
Alhassid Y, G\"ursey F and Iachello F 1983 {\em Phys. Rev. Lett.} {\bf 50}
873--6
\bibitem{BB93}
Braibant S and Brihaye Y 1993 {\em J. Math. Phys.} {\bf 34} 2107--14
\bibitem{KLS94}
Kofman L, Linde A and Starobinsky A A 1994 {\em Phys. Rev. Lett.} {\bf 73} 3195--8
\bibitem{GKLS97}
Greene P B, Kofman L, Linde A and Starobinsky A A 1997 {\em Phys. Rev. D} {\bf 56}
6175--92
\bibitem{Tu89}
Turbiner A V 1989 {\em J. Phys. A: Math. Gen.} {\bf 22} L1--L3
\bibitem{GKOqes94}
Gonz\'alez-L\'opez A, Kamran N and Olver P J 1994 {\em Contemporary Mathematics}
{\bf 160} 113--40
\bibitem{Us94}
Ushveridze A G 1994 {\em Quasi-Exactly Solvable Models in Quantum Mechanics}
(Bristol: IOP)
\bibitem{Ol97}
Olver P J 1997 {\em GROUP21: Physical Applications and Mathematical Aspects
of Geometry, Groups, and Algebras}, ed H-D Doebner {\em et al.} (Singapore: World
Scientific) 285--95
\bibitem{Ra80}
Razavy M 1980 {\em Am. J. Phys.} {\bf 48} 285-8
\bibitem{FGR99}
Finkel F, Gonz\'alez-L\'opez A, and Rodr\'\i guez M A 1999
{\em Preprint} math-ph/9905020
\bibitem{Wa87}
Ward R S 1987 {\em J. Phys. A: Math. Gen.} {\bf 20} 2679--83
\bibitem{UZ92}
Ulyanov V V and Zaslavskii O B 1992 {\em Phys. Rep.} {\bf 216} 180--251
\bibitem{BG93}
Brihaye Y and Godart M 1993 {\em J. Math. Phys.} {\bf 34} 5283--91
\bibitem{Tu88} Turbiner A V 1988 {\em Commun. Math. Phys.} {\bf 118} 467--74
\bibitem{GKOnorm93}
Gonz\'alez-L\'opez A, Kamran N and Olver P J 1993
{\em Commun. Math. Phys.} {\bf 153} 117--46
\bibitem{Tu92} Turbiner A V 1992 {\em J. Phys. A: Math. Gen.} {\bf 25} L1087--L93
\bibitem{FK98} Finkel F and Kamran N 1998
{\em Adv. Appl. Math.} {\bf 20} 300--22
\bibitem{TU87} Turbiner A V and  Ushveridze A G 1987 {\em Phys. Lett.} {\bf
A126} 181--3
\bibitem{FGRorth96}
Finkel F, Gonz\'alez-L\'opez A and  Rodr\'\i guez M A 1996
{\em J. Math. Phys.} {\bf 37} 3954--72
\end{thebibliography}
\end{document}